\documentclass{ws-mplb}
\usepackage{amsfonts,amssymb,amsmath}
\usepackage{graphicx}
\usepackage[usenames, dvipsnames]{color}
\usepackage[super,sort,compress]{cite} % sort => 1,2,3,5; nosort => 1,3,2,5; compress => 1-3,5; nocompress => 1,2,3,5
\usepackage{adjustbox}
\begin{document}

\markboth{Authors' Names}{Instructions for typing manuscripts (paper's title)}

\catchline{}{}{}{}{}
%
%%%%%%%%%%%%%%%%%%%%%%%%%%%%%%%%%%%%%%%%%%%%%%%%%%%%%%%%%%%%%%%%%%%%

\title{Effect of correlation on the traffic capacity of Time Varying Communication Network}

\author{\footnotesize Suchi Kumari}

\address{Department of Computer Science Engineering \\ National Institute of Technology Delhi, India\\
suchisingh@nitdelhi.ac.in}

\author{Anurag Singh}

\address{Department of Computer Science Engineering \\ National Institute of Technology Delhi, India\\
anuragsg@nitdelhi.ac.in}

\maketitle

\begin{history}
\received{(Day Month Year)}
\revised{(Day Month Year)}
\end{history}

\begin{abstract}
The network topology and the routing strategy are major factors to affect the traffic dynamics of the network. In this work, we aim to design an optimal time-varying network structure and an efficient route is allocated to each user in the network. The network topology is designed by considering addition, removal, and rewiring of links. At each time instants, a new node connects with an existing node based on the degree and correlation with its neighbor. Traffic congestion is handled by rewiring of some congested links along with the removal of the anti-preferential and correlated links. Centrality plays an important role to find the most important node in the network. The more a node is central, the more it can be used for the shortest route of the user pairs and it can be congested due to a large number of data coming from its neighborhood. Therefore, routes of the users are selected such that the sum of the centrality of the nodes appearing in the user's route is minimum. Thereafter, we analyze the network structure by using various network properties such as the clustering coefficient, centrality, average shortest path, rich club coefficient, average packet travel time and order parameter.
\end{abstract}

\section{Introduction}
The structure and dynamics of complex networks attracted much attention from the researchers of different areas in recent years. It has been widely accepted that the topology and degree distribution of networks have intense effects on the process dynamics on these networks, including disease spreading \cite{Holme2016}, information diffusion \cite{guille2013information}, traffic movement \cite{2016congestion,jiang2013,onset}. Due to the increasing traffic volumes on the networks e.g., road networks, social and data communication networks, fulfillment of user's demand and minimization of traffic congestion is a challenging task. The performance of the data communication networks strongly depends on its data forwarding capacity which is determined by the structure of the underlying network. In this context, an optimal time-varying communication network model is designed to avoid congestion in the network and user's route is selected based on centrality information especially, betweenness centrality.

It is found that the communication networks are scale-free (SF) \cite{bara} and are more susceptible to traffic congestion than some homogeneous networks \cite{erdos1960}. In SF networks large degree nodes posses a large volume of data hence, congestion usually starts at these nodes and then spreads to the whole network. Therefore, researchers proposed various strategies \cite{chen2012}, which can be classified into \textbf{hard} and \textbf{soft} strategies in order to handle traffic congestion and enhance network capacity. The restructuring of network topology comes under hard strategies. Zhao \textit{et al.} redistributed a load of heavily loaded nodes to others \cite{onset}, some connections are removed between large degree nodes \cite{liu}, high betweenness centrality nodes are removed first \cite{zhang} and links are added between the nodes with long distance \cite{huang2010}. Jiang \textit{et al.} \cite{jiang2013optimal}, assigned capacity dynamically to each link proportional to the queue length of the link. Some fraction of links is rewired based on node's degree information and betweenness centrality \cite{jiang2013}.  Chen \textit{et al.}\cite{chen2018} rewired the link against traffic congestion and proved that the network should have a core-periphery structure. 

Sometimes it is impossible to modify the network topological structure and it also incurs a high cost to change the structure of the network. Hence, a soft strategy based on finding a better routing strategy is preferable to enhance the network capacity. Yin \textit{et al.} \cite{yin2006traffic} chose an efficient path (EP) for routing. Zhao \textit{et al.} \cite{zhang2017} assigned different routes with different traffic flow priorities and it is shown that traffic capacity is enhanced by approx $ 12\% $ compared with the initial EP approach. In communication networks, two rates are associated with each node: packet generation rate, $ \lambda $ and packet forwarding (delivery) rate, $ C $. Tang \textit{et al.}\cite{tang2009} considered the $ \lambda $ as a periodic function of time and proposed a mixed routing strategy to enhance transportation efficiency.  For small $ \lambda $, the shortest paths are used to deliver the packets and when $ \lambda $ is large, an efficient routing method is used and loads are redistributed from central nodes to others.  The capacity of the network is maximized using a routing strategy based on the minimum information path and average routing centrality degree of the node is calculated to analyze the traffic load on nodes of different degrees \cite{wang2011}. Some of the researchers studied transport processes on multilayer \cite{du2016} or multiplex networks \cite{2016congestion} with emphasis on optimizing the network capacity and transmission efficiency. Yang \textit{et al.}\cite{yang2017} established a relationship between traffic congestion and network lifetime for the network with moving nodes in a defined area and it is concluded that the network lifetime is inversely related to traffic congestion.

The previous researches demonstrate the underlying network topologies and routing approaches have significant impacts on the overall network performance. The strategies for optimizing network topology can be divided into two parts: restorative and proactive strategies. Restorative strategies include methods to do some changes in existing networks such that congestion is minimized \cite{liu,zhang,huang2010,jiang2013,chen2018}. Through proactive strategies, new links are added in such a way that the optimal network structure is formed\cite{jiang2013optimal,zhao2007}. A proactive strategy like \cite{zhao2007} converts the network into the homogeneous network but most of the real-world networks are scale-free networks. Therefore, the problem statement for the proposed work may be categorize under following categories. (i) Design an optimal network structure by considering both the proactive and the restorative strategy. A time-varying communication network (TVCN) model is proposed in such a way that the addition of new links are based on a proactive strategy while removal and rewiring of links follow restorative strategy. The probability that the new node will be attached to the node in the existing network is proportional to the degree of the existing node and inversely proportional to the correlation of the existing node with its neighbor. Few links of the congested nodes are rewired and connected with the nodes with preferential attachment and having less correlation with its neighbor. Some correlated and anti-preferential links are also removed from the network. (ii) The novelty of the proposed network structure is checked by studying different network parameters such as the clustering coefficient ($ C_lC $), centrality, average path length (APL), and rich club (RC) coefficient. (iii) The route of each user is selected with the help of centrality of the node in the network. The betweenness centrality (BC) is used to measure the extent to which a node lies on shortest paths between other node pairs. If a node is more betweenness central then it may appear in a large number of users' route. The nodes with maximum BC values are the most congested nodes in the network \cite{onset}. Therefore, routes of the users are selected such that the sum of the BC  of the nodes appearing in the user's route is minimum. As a result, it is found that the path constructed through small betweenness central nodes is least congested than the other shortest paths. Simulation results show that the proposed routing strategy effectively enhances the transmission capacity and reduce the load of the networks.

In Section 2 some existing network models and proposed network model are described. Section 3 discusses about the traffic flow models. Section 4 presents the simulation results, and in Section 5, conclusions and future research plan are discussed.
\vspace*{-0.5 cm}
\section{Network Models}
The concept of evolving network is given by Barabasi-Albert \cite{bara} where, a new node attaches with an existing node through preferential attachment. In BA model, only addition of nodes and links are considered and a new link will appear only when new node arrives. But, in real scenario, links may appear or disappear at any time. Hence, two models are discussed: time varying communication network (TVCN) model and disassortative time varying communication network (DTVCN) model. 
\subsection{Time Varying Communication Network Model}
Time varying communication network (TVCN) model can be represented as $ G = (N, E, \tau) $, where, $ N $ is the set of nodes, $ \tau $ is the set of time instants for which the TVCN is defined, and $ E \subseteq N \times \tau \times N \times \tau $ is the
set of links. A dynamic link, $ e_{ab} $ between source, $ a $ and destination, $ b $ in a TVCN, $ G $ is defined as an ordered quadruple $ e_{ab} = (a, t_i, b, t_j)$, where $ a, b \in N(G) $ while $ t_i, t_j \in \tau(G) $ are the source and destination time instants, respectively. In this paper, two types of dynamic links are considered: temporal link and spatial link. Temporal link is used to connect the same node at two distinct time instants, $ e_{aa} $ is in the form of $e_{aa} = (a, t_i, a, t_j) $, where $ t_i \neq t_j $. Spatial links connect two different nodes at the same time instant, $ e_{ab} $ is in the form of $e_{ab} = (a, t_i, b, t_i) $, where $ a \neq b $. 
\begin{figure}[!htb]
\begin{center}
\includegraphics[width=\linewidth, height=1.5 in]{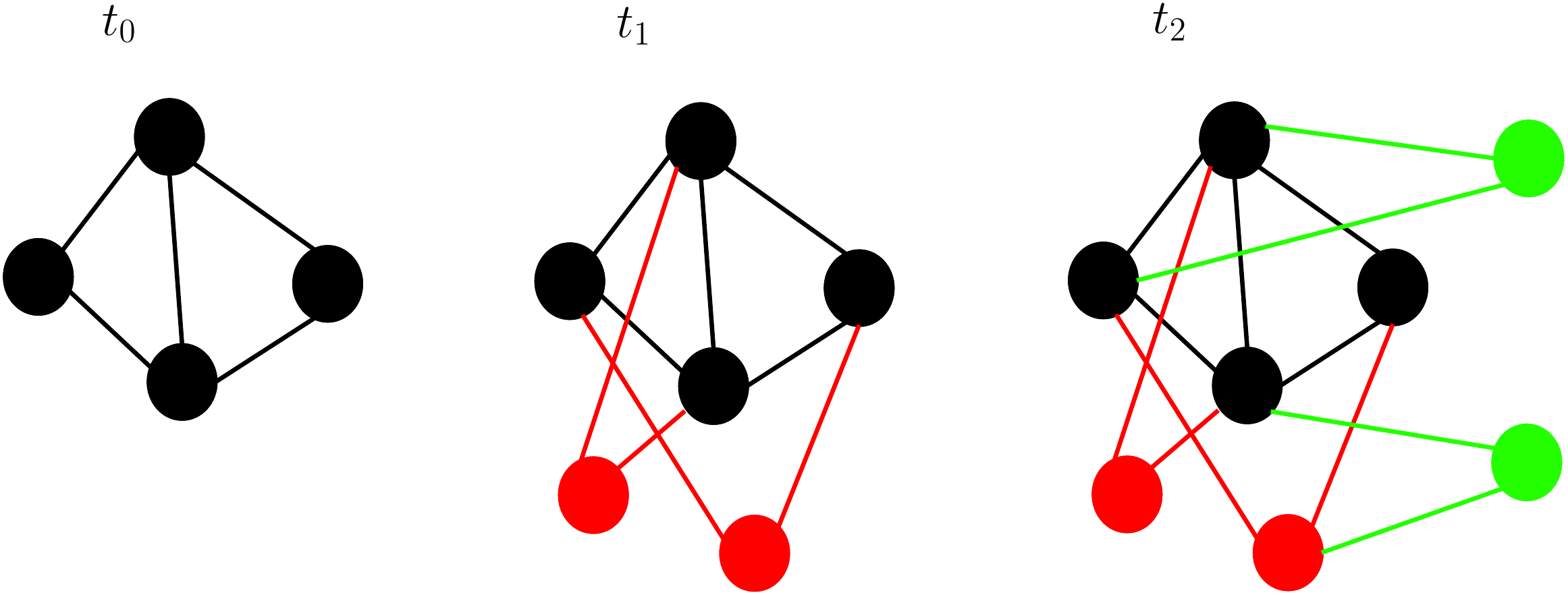}
\caption{A Time Varying Communication network at $ t_0, t_1 $ and $ t_2 $ time instants.}
\label{f1}
\end{center} 
\end{figure}

In Fig. \ref{f1}, a network at different time instances is shown. At each time instant, few nodes appear and some links are also getting added into the network. A TVCN model is proposed by us \cite{kumari2017} where, three operations are performed at each time instants; addition of nodes and links, removal and rewiring of links. Addition and rewiring of links are based on the preferential attachment where, probability $ \Pi $ that a node  $ i $ will be selected through preferential attachment is proportional to its degree and is given by, $ \Pi = \frac{k_i}{\sum_{j \in N} k_j} $. Removal of links are based on anti preferential attachment and the probability $ \Pi' $ of selecting node $ i $ with anti-preferential attachment is given by, $ \Pi' = \frac{1}{|N|-1} \left(1-\frac{k_i}{\sum_{j \in N} k_j}\right) $.  At each time instant $ t_i \in \tau $, a new node $ i $ is added to the network (expansion) and a number $ M (\leq n_0) $ is selected for network dynamics.  Total $ \vartheta M$ links are added from the new node and $ (1-\vartheta)M $ links are used for alteration (rewiring and removal of links) in the network; $ 0 < \vartheta < 1 $.

The scale-free (BA)model and TVCN model assume that the new node will prefer to attach with the nodes in the existing network based on the value of the degree of the existing node. But, in most of the networks, this assumption may not be true, as nodes can't acquire links unconditionally. There is a limit on packet forwarding rate i.e., the capacity of the node. If more links are attached to a node then, it may happen that the node will be a part of a large number of user's shortest paths and leads to congestion in the network. Therefore, we modified the probability to account for the congestion and introduced degree-degree correlation (DDC) as a multiplicative factor to the preferential attachment probability. DDC is a network property in which nodes with similar attributes, such as degree, tend to be connected. DDC is used to divide networks into three types: assortative, disassortative and neutral networks \cite{newmanassortative}. For assortative networks, hubs(small degree nodes) tend to link to other hubs (small degree nodes). In a disassortative network, hubs (small degree nodes) avoid each other, linking instead to small-degree nodes (hubs). While in the neutral network, the number of links between nodes is random. 

\subsection{Disassortative Time Varying Communication Network Model }
%\vspace*{-.4cm}
In the network, a node $ i $ may generate packets with a packet generation rate, $ \lambda_i $ and may forward packets according to its capacity, $ C_i $. For the smaller value of the packet generation rate, $ \lambda $ system remains in free flow state as every packet is getting delivered. But with the increasing value of $ \lambda $, a point is reached where system converts into congested phase and this point of phase transition is known as critical packet generation rate, $ \lambda_c $. The value of $ \lambda_c $ is affected by the topology of the network.  DDC has an important influence on the structural properties of the network and is one of the deciding parameters to find congestion in the network \cite{2012theoretical} as well. Communication network comes under disassortative network. For that reason, in this paper, a disassortative TVCN (DTVCN) model is proposed to achieve a higher value of $ \lambda_c $.  A new node may be attached to the existing nodes by preferring higher degree and disassortativity with the neighbors of the existing nodes. Some nodes rewire their links and attach with the nodes by preferring degree and disassortativity with the nodes' neighbors. While some fraction of anti-preferential and correlated links are removed from the network. In this way, congestion is minimized and we get the higher value of $ \lambda_c $.

A new node, $ i $  will establish a new connection with a node $ v $ by preferring the higher degree of the node $ v $ and normalized disassortativeness, $ \zeta_v $ of node $ v $ with its neighbors, $ Ne(v) $ in the network. The value of $ \zeta_v $ is dependent on the correlation $ r_{deg}(v) $ of node $ v $ with its neighbors. The $ r_{deg}(v) $ measures maximum disassortativeness of a node $ v $ with its neighbor. If a node $ v $ is more disassortative to its neighbor then the selection probability will be increased. The probability $ \Pi_v^r $ of node $ i $ to attach with the node $ v $ may be defined as,
\begin{equation}
\Pi_v^r =  \frac{k_v}{\sum_u k_u}  \zeta_v\label{e5}
\end{equation}
Where, $ \zeta_v = \frac{r_{deg}(v)}{\sum_{n=1}^{|Ne(v)|} r_{deg}(v,n)} $, $ r_{deg}(v) = min \mbox{ } r_{deg}(v,n), \forall {n: n \in Ne(v)} $, and the value of $ r_{deg}(v,n) $ is scaled from the range of $ [-1,1] $ to $ [0,2] $.
In the rewiring strategy, the link $ e_{jk} $ between nodes $ j $ and $ k $ is removed if the node $ j $ shows assortativeness $ 0 < r_{deg} \leq 1 $ with node $ k $. The node $ j $ will rewire its connection and connects with a node $ u $ with probability $ \Pi_u^r $. The probability $ \Pi_v^{r'} $ of selecting a node $ v $ with anti-preferential attachment and high correlation with its neighbors is given by, $$ \Pi_v^{r'} = \left( 1-\frac{k_v}{\sum_j k_j}\right) (1-\zeta_v) $$.

\section{Traffic Model}
Traffic flow model is based on different routing algorithms in communication networks. As the topologies of various networks are different hence, data delivery capacity of each node is considered as different for different nodes, depending on the effect of the node on the other nodes in the network. Some nodes are endorsed by the nodes which are already congested and cause an increase in load at that node. The aim of the proposed approach is to assign more capacities to the congested node in order to reduce data traffic in the network. As Eigenvector centrality is used to define the impact of the node on the the other nodes in the network and is considered in the computation of capacity of the node. Therefore, data forwarding capacity, $ C_i $ of node $ i $ may be defined as,
\begin{equation}
C_i = \beta \hat{x}(v) |N|.\label{e1}
\end{equation}
Where, $ \hat{x}(v) $ is Eigenvector centrality of node $ v $, $ \hat{x}(v) = \frac{1}{\kappa} \sum_{j \in N} a(vj) \hat{x}(j) $ and $ \kappa $ is a constant. The term, $ \beta $ is a considerably modest fractional value and is a controlling parameter for capacity of the nodes. Packets are forwarded through the shortest paths from the source node to the destination node into the network. Therefore, the probability to pass through a node $ i $ is provided as $ \frac{g(i)}{\sum_{j= 1}^{|N|} g(j) } $. At each time step, average number of packets generated is $ \alpha | N | $. Hence, the probability of node $ i $ to generate a packet, $ \lambda_i $,
\begin{equation}
\lambda_i = \alpha \mathcal{D} | N | \frac{g(i)}{\sum_{j= 1}^{|N|} g(j)}. \label{e2}
\end{equation}
Where $ \mathcal{D} $ is the diameter of the network. The term, $\alpha$ is a small fractional value and is a controlling parameter for $ \lambda_i $ for the node $ i $. The capacity of a node increases with increase in the value of $ \beta $. The sum of the capacities $ C_i $s of each node $ i $ is known as the capacity $ C $ of the network. Similarly, the sum of the packet generation rates, $ \lambda_i $s of each node $ i $  is termed as the traffic load, $ \lambda $, of the network. If traffic load exceeds the traffic capacity of the network then the system will be in the congested state otherwise, it will remain in the free flow state. The point at which the phase transition occurs from free flow state to congested flow state is known as the critical point as well as the critical packet generation rate ($ \lambda_c $). There are three possible relationships between $ \lambda $ and $ C $: (i.) $ \lambda < C $ implies the system in free flow state, (ii.) $ \lambda = C $ shows the boundary case for congestion and (iii.) $ \lambda > C $ allows a system in congestion.  Local congestion of a node $ i $ is calculated according to its $ C_i $ and $ \lambda_i $. If $ \lambda_i > C_i$ of node $ i $ then the node will be congested and these nodes increase overall network congestion. 

The node with the maximum BC can be easily congested hence, it is necessary to consider only the traffic load of this node. The maximum number of packets that can be forwarded through node $ max $ is $ C_{max} $ and the total number of packets accumulated at maximum BC node is $ \lambda_c \mathcal{D} | N | \frac{g({max})}{\sum_{j= 1}^{|N|} g(j) }  $. The value of $ \alpha $ in Eq. \eqref{e2} is replaced by $ \lambda_c $ and theoretical estimation of critical packet generation rate is given by $ \lambda_c = \frac{C_{max} (|N|-1)}{g({max})} $ \cite{onset}. The nodes with maximum packet forwarding capacity and maximum BC are denoted by $ C_{max} $ and $ g(max) $, respectively. The average number of packets for a packet generation rate ($ \lambda $) over a time window $ \Delta t $ can be calculated by using an order parameter $ \theta(\lambda) $ and is given by,
\begin{equation*}
 \theta(\lambda) =  lim_{t \rightarrow \infty} \frac{C}{\lambda}\frac{\langle \Delta \mathbb{P}  \rangle}{ \Delta t}
\end{equation*} 
$ \Delta \mathbb{P} = \mathbb{P}(t + \Delta t)-\mathbb{P}(t) $, $ \mathbb{P}(t) $ is number of packets at time $ t $ and $ \langle . \rangle $ shows the average value over the time window $ \Delta t $. $ \langle \Delta \mathbb{P}  \rangle = 0 $ indicating that there is no packet in the network for $ \theta = 0 $, system is in free flow state. The ratio of $ C $ and $ \lambda $ provides the quantity of undelivered packet over $  \Delta t $ in the network.

The communication network is shared by a large number of users. But the user's information is not available at the time of design of network structure. If a node appears in the path of multiple users then the capacity of the node will be divided equally among the users. Let us consider a set of $ R $ users who want to access the network and each user $ r $ selects its source node and destination node randomly. There exist multiple shortest paths $ \sigma_u^r (s \rightarrow d) $ for each user $ r $,  from the source node $ s $ to the destination node $ d $ for $ u \in [1, \mathcal{K}] $ where $ \mathcal{K} $ is the number of the SPs ($ \sigma_u $). Since the value of $ \lambda_c $ of the network is affected by the maximum BC node and sometimes, it may happen that maximum BC node appears in user's route which leads to congestion in the route. Hence, the probability of selecting the shortest path $ \sigma_u^r(s \rightarrow d) $ depends on the value of sum of BC ($ W_g(\sigma_u^r ( s \rightarrow d)) $) of the nodes appearing in the route of user $ r $ and may be defined as,
\begin{equation}
 W_g(\sigma_u^r ( s \rightarrow d)) =\sum_{n: n \in \sigma_u^r(s \rightarrow d)} g(n)
\end{equation}
After calculating  $ W_g(\sigma_u^r ( s \rightarrow d) $, we assign weight on each SP from  $ s $ to $ d $. The SP, $ \sigma^r $ with minimum value of $ W_g(\sigma_u^r ( s \rightarrow d) $ is chosen for routing of the data through the network and it can be represented as,
\begin{equation}
W_g \mbox{ (min) } = \min_u\mbox{  } W_g(\sigma_u^r ( s \rightarrow d))
\end{equation}
Similarly, the SP with maximum value of $ W_g(\sigma_u^r ( s \rightarrow d) $ can be denoted as $ W_g $(max).

\begin{figure}[!htb]
\begin{center}
\includegraphics[width=0.75\linewidth, height=2.5 in]{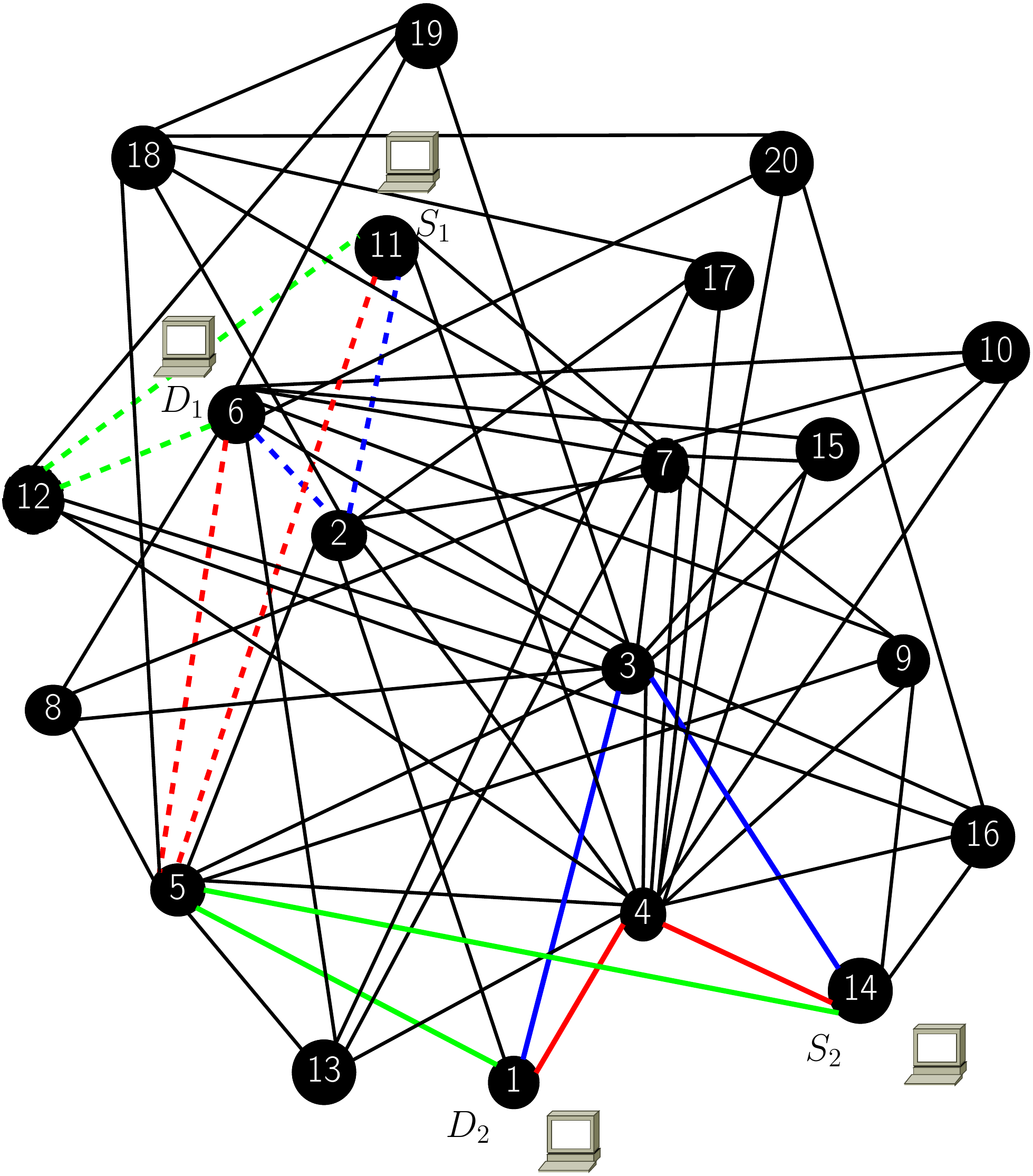} 
\caption{A scale-free network generated through the BA model with $ |N| = 20 $ and $\langle k \rangle = 4 $. Red, blue and green color solid line shows route of User1 and dashed line is for the route of User2 under different routing strategies; $ W_g $(max), a random shortest path (SP) and $ W_g $(min) respectively.}
\label{f6}
\end{center} 
\end{figure}
The topological structure of the BA model is shown in Fig. \ref{f2}. Different routing strategies ($ W_g $(min), $ W_g $(max) and a random shortest path (SP)) are applied for assigning shortest paths to each user. The sum of BC, $ W_g $, of the nodes appearing in the shortest paths for different routing strategies is computed. In Fig. \ref{f2}, two users, User1, and User2, want to access the network. The value of $ W_g $, for the SP through red, blue and green color solid lines, is $ 0.2020, 0.1197 $ and $ 0.0756 $ respectively. The SP is chosen through red, blue and green color dotted line for User2 gives the value of $ W_g $ as $ 0.0756, 0.0350 $ and $ 0.0198 $ respectively. 

At a particular time instant, only  $ R $ users want to access the network and few nodes will be their point of interest. Therefore, instead of considering the whole network to calculate $ \lambda_c $, a subnetwork consisting all the nodes appear in the users' route is only considered. As the size of the network increases, packets are more likely to be routed to the nodes with higher BC and packets are more likely to be accumulated at these nodes, resulting in traffic congestion. But, the proposed routing strategy avoids larger BC nodes hence, we may get a higher value of $ \lambda_c $ and there will be less congestion on the nodes those are the part of the subnetwork. 

In a network, if a packet is generated then it needs to be delivered. Once the packet is reached to its destination it is removed from the network. The average time $ \langle T \rangle $ to deliver all the packets to the destination nodes are dependent on the route of the user. Each node $ i $ generates packets with rate $ \lambda_i $. If $ \lambda_i < C_i $ then all the packets will be forwarded to the next node towards its destination otherwise it needs to wait at the end of the queue. After that, packets will be processed on a first come first serve basis. From this, we can infer that the packets waiting in the queue increase $ \langle T \rangle  $ and $ \theta $ as well. The average packet travel time can be formulated as,
\begin{equation}
\langle T \rangle = \sum_{r:r \in R} \sum_{n: n \in \sigma^r} \lambda_n/ \min_{n: n \in \sigma^r}C_n
\end{equation}

\section{Simulation and Result}
For the dynamics of DTVCN model, the simulation starts by establishing the infrastructure of the network. In this paper, the parameters are set with the value as seed node $ n_0 = 5 $, number $ M (\leq t) $, fraction of newly added links $ \vartheta $ range in $ (0,1) $, fraction of rewired links $ \gamma $ is in the range of $ (0.5, 1) $, with network size ranging from $ |N| = 10^2 $ to $ |N| = 5 \times 10^3 $. Any node can be included in the user's $ (s,d) $ sets or may participate in routing also. The capacity of each node is proportional to the Eigen vector centrality of the node. At each time stamp, the degree of the nodes will be different, hence, data forwarding capacity as well as data generation rate change accordingly. Here, the range of DDC is scaled from $ (-1,1) $ to $ (0,2) $. The value of the controlling parameters $ \alpha $ and $ \beta $ may be any fractional value.

\begin{figure}[!htb]
\begin{center}
\includegraphics[width=0.5\linewidth, height=2 in]{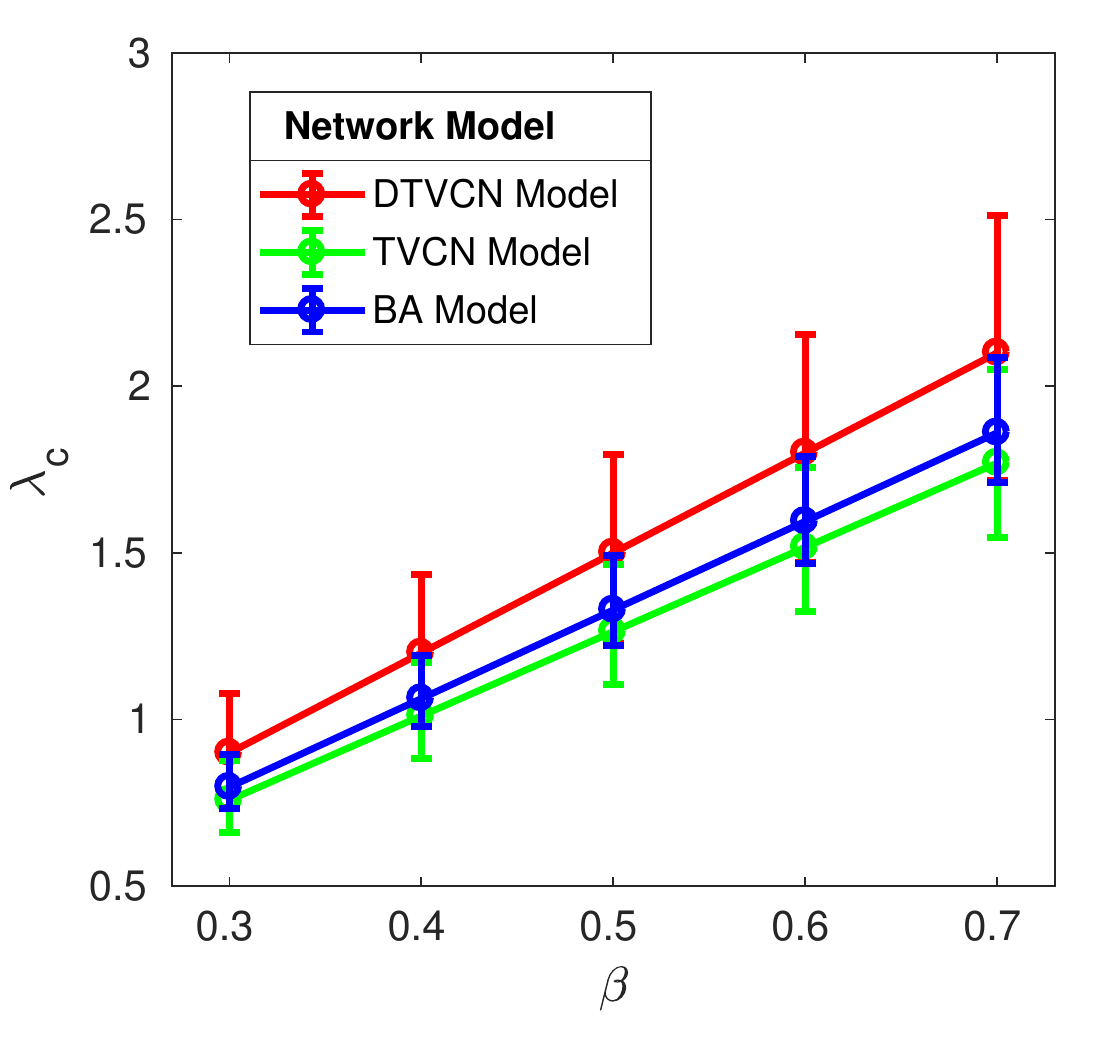} 
\caption{The value of $ \lambda_c $ for different values of $ \beta $. Each result value is the average of $ 10 $ independent realizations of BA model, TVCN model and DTVCN model.}
\label{f2}
\end{center} 
\end{figure}

In Fig. \ref{f2}, critical packet generation rate, $ \lambda_c $ is evaluated for the networks designed through different strategies.  $ \beta $ is the controlling parameter for capacity $ C $. As $ \lambda_c $ is proportional to the capacity of the maximum BC node hence, it increases with $ beta $.  DTVCN model considers congestion at the time of topology design hence, it gives the higher value of $ \lambda_c $. In the TVCN model, addition and rewiring strategies are based on preferential attachment and it will increase the degree of higher degree nodes which leads to increased congestion in the network. Therefore, the model gives least value of $ \lambda_c $ for different $ \beta $. 

\begin{figure}[!htb]
\begin{center}
\includegraphics[width=\linewidth, height=3.5 in]{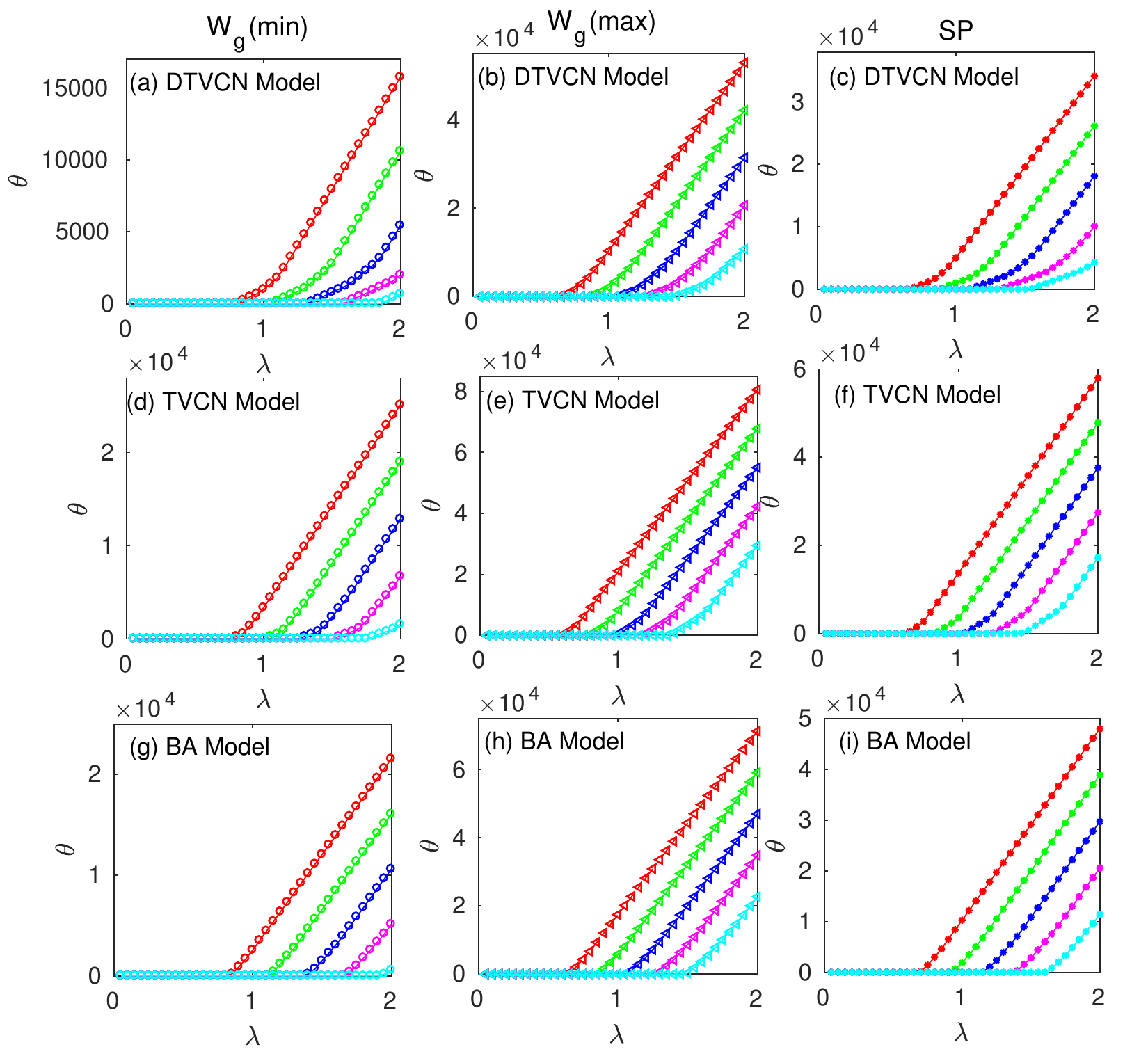} 
\caption{The order parameter $ \theta $ versus the packet generating rate $ \lambda $ under different routing strategies ($W_g $(min), $ W_g $ (max) and  Shortest Path (SP)) and the network designed through DTVCN model, TVCN model and BA model. Red, green, blue, magenta and cyan color blocks correspond to the simulations of $ \beta = 0.3, 0.4, 0.5, 0.6, 0.7 $ respectively. Each result value is the average of $ 10 $ independent realizations.}
\label{f3}
\end{center} 
\end{figure}

Network topology and routing strategies both are deciding parameters for the load at the network. If we increase the flow of packets through the congested node then, the total number of undelivered packets will increase accordingly. Multiple shortest path exists for the different source node and the destination node. If the packets are sent through the path with $ W_g $(min) then the total number of accumulated packets $  \theta $ is lower for distinct values of $ \lambda $. The random SP offers the value of $ \theta $ in between $ W_g $(max) and $ W_g $(min). The network designed through DTVCN model and routes selected through $ W_g $(min) give least value of $ \theta $ for different $ \lambda $ and the network designed through TVCN model and the route selected through $ W_g $(max) gives highest value of $ \theta $. As a result, it is inferred that the network designed through DTVCN model outperforms than other models for all routing strategies. 

\begin{figure}[!htb]
\begin{center}
\includegraphics[width=\linewidth, height=3.5 in]{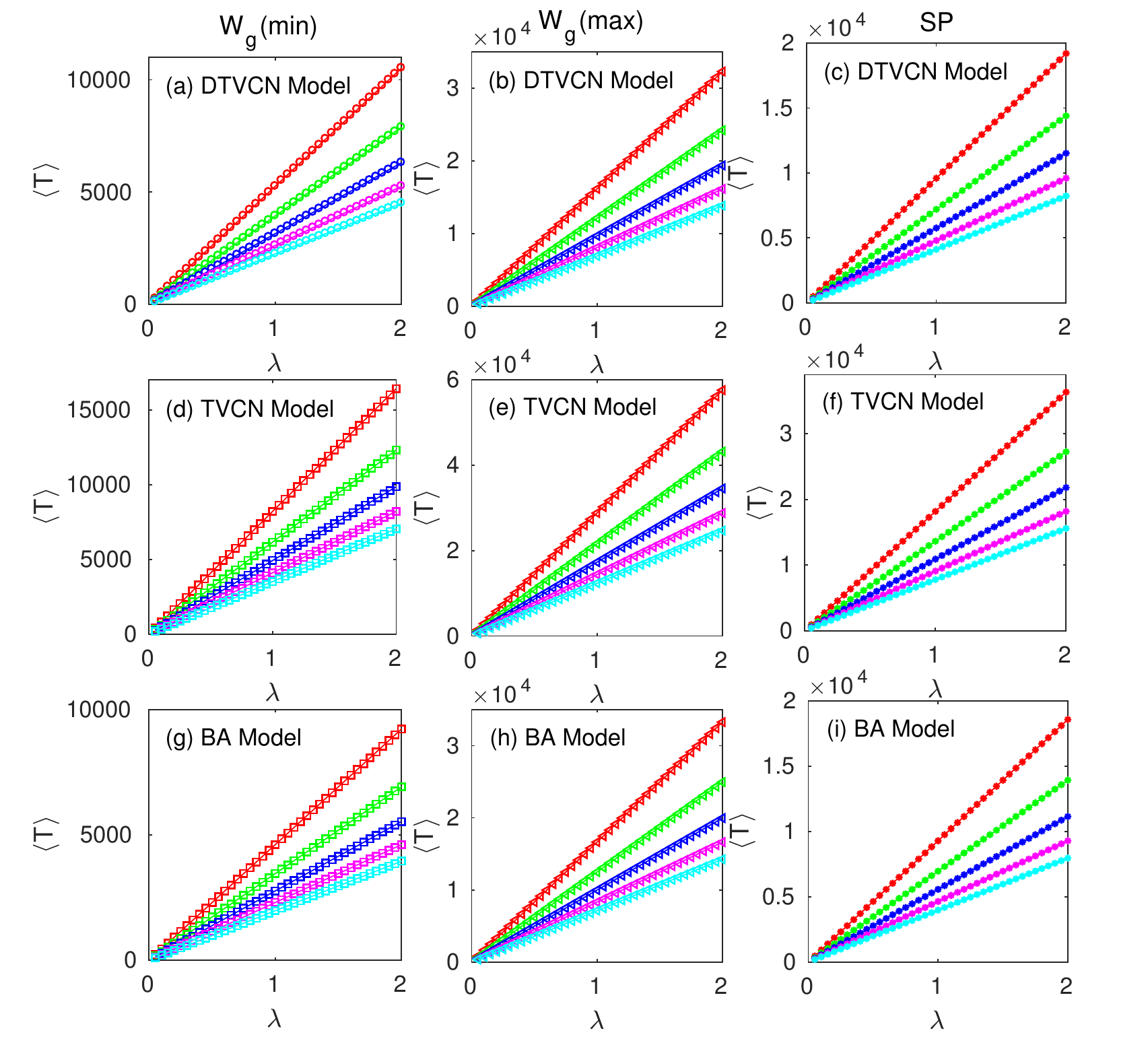} 
\caption{Average packet travel time, $ \langle T \rangle $ versus packet generating rate $ \lambda $ for different value of controlling parameter, $ \beta $ under different routing strategies ($W_g $(min), $ W_g $ (max) and  Shortest Path (SP)) and the network designed through DTVCN model, TVCN model and BA model. Red, green, blue, magenta and cyan color square blocks correspond to the simulations of $ \beta = 0.3, 0.4, 0.5, 0.6, 0.7 $ respectively. Each result value is the average of $ 10 $ independent realizations.}
\label{f4}
\end{center} 
\end{figure}

Congestion increases the number of the accumulated packet in the route of the user and the average packet travel time, $ \langle T \rangle $. In Fig. \ref{f4}, $ \langle T \rangle $ is maximum when the routing strategy, $ W_g $(max) is applied on the network designed through all the models (DTVCN model, TVCN model and BA model) while $ \langle T \rangle $ is minimum for $ W_g $(min) routing approach. The value of $ \langle T \rangle $ for the random shortest path lies in between $ W_g $(min) and $ W_g $(max) routing approaches. The performance of the network designed through TVCN model for all routing approaches is lower than the other models. $ \langle T \rangle $ for BA model and DTVCN model are approximately same for their respective routing strategies. 

\begin{figure}[!htb]
\begin{center}
\includegraphics[width=\linewidth, height=1.75 in]{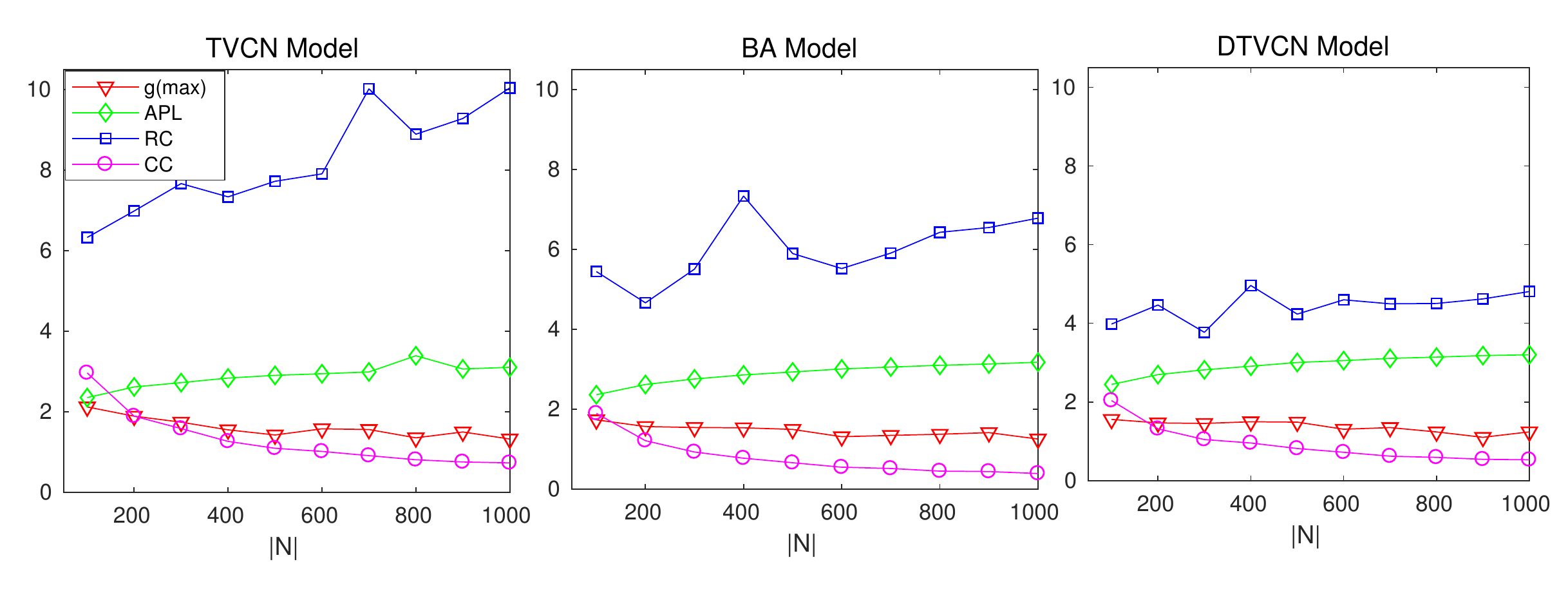} 
\caption{Network size $ |N| $, $ g(max) $, average path length (APL), rich club coefficient ($ RC $) and clustering coefficient ($ C_lC $) for all the models; DTVCN, TVCN and BA when, the network size, $ |N| $ varying from $  10^2 $ to $ 10^3 $. Each result value is the average of $ 20 $ independent realizations}
\label{f5}
\end{center} 
\end{figure}
 
In Fig. \ref{f5}, various network properties such as maximum betweenness centrality ($ g(max) $), average path length (APL), rich club coefficient ($ RC $) and clustering coefficient ($ C_lC $) are studied. All these are the properties of the network as a whole and not just of the individual node. This allows for the analysis of how the whole network changes and not just the structure around some particular node. Rich club phenomenon is characterized when the hubs are on average more intensely interconnected than the nodes with smaller degrees. When the nodes are with a large degree than $ k $ tends to be more connected than the nodes with the smaller degree. The rich club phenomenon refers to the tendency of hub nodes to connect with other higher degree nodes than the nodes with a smaller degree. Presence of $ RC $ increases load and congestion on the connecting link between two hubs. Most of the users want to send data through shortest paths i.e., through hub nodes and congestion at hub nodes will reduce $ C $ and efficiency of the networks. The proposed DTVCN model takes care of congestion and offers the lowest value of RC than the other two models; the BA model and TVCN model. The value of $ g(max) $ for DTVCN model is minimum and TVCN model is maximum for different value of $ |N| $. The average path length (APL) of the network structured through all the models lie between $ 2 $ to $ 3 $. The clustering coefficient, $C_lC$ decreases with increasing value of network size, $ |N| $ and the $ C_lC$ of BA model is minimum and TVCN model is maximum. The length of the average shortest path of all the models increases with increasing value of the $ | N | $. The quantitative value of the network properties is shown in Table \ref{tab1}. 

\begin{table}[!htb]
\caption{Network size $ |N| $, $ g(max) $, average path length (APL), rich club coefficient ($ RC $) and clustering coefficient ($ C_lC $) for all the models; DTVCN, TVCN and BA when, the network size, $ |N| $ varying from $  10^2 $ to $ 10^3 $.} % title name of the table
 % centering table
 \center
 \adjustbox{max height=\dimexpr\textheight-4.5cm\relax,
           max width=0.7\textwidth}{
\begin{tabular}[1]{|l| c| c| c| c| c|}
\hline\hline
 & Network & $ g(max) $ & Average &  $ RC $ & Clustering  \\ [0.5ex]
$ |N| $ & Model &  & Path Length & & Coefficient \\
\hline\hline
 & DTVCN & $ 0.1466 $ & $ 2.6938 $ & $ 4.4639 $ & $ 0.1316 $  \\[0.25ex]
$ 200 $ &  TVCN & $ 0.1890 $ & $ 2.6154 $ & $ 6.9843 $ & $ 0.1890 $  \\[0.25ex]
& BA & $ 0.1562 $ & $ 2.6140 $ & $  4.6577 $ & $ 0.1208 $   \\[0.25ex]
\hline
 &  DTVCN & $ 0.1495 $ & $ 2.9076 $ & $ 4.9605 $ & $ 0.0955 $  \\[0.25ex]
$ 400 $ & TVCN & $ 0.1551$ & $ 2.8362 $ & $ 7.3350 $ & $ 0.1264$  \\[0.25ex]
 & BA & $ 0.1530 $ & $ 2.8538 $ & $ 7.3302 $ & $ 0.0770 $  \\[0.25ex]
\hline
 &  DTVCN & $ 0.1305 $ & $ 3.0502 $ & $ 4.5967 $ & $ 0.0719 $  \\[0.25ex]
$ 600 $ & TVCN & $ 0.1576 $ & $ 2.9438 $ & $ 7.9056 $ & $ 0.1015 $  \\[0.25ex]
 & BA & $ 0.1307 $ & $ 3.0027 $ & $ 5.5156 $ & $ 0.0544$  \\[0.25ex]
\hline
 &  DTVCN & $ 0.1237 $ & $ 3.1391 $ & $ 4.5039 $ & $ 0.0589 $ \\[0.25ex]
$ 800 $ & TVCN & $ 0.1352 $ & $ 3.3902 $ & $ 8.8899 $ & $ 0.0808 $ \\[0.25ex]
 & BA & $ 0.1369 $ & $ 3.0942 $ & $ 6.4224 $ & $ 0.0446 $ \\[0.25ex]
\hline
 &  DTVCN & $  0.1238 $ & $ 3.1983 $ & $ 4.8072 $ & $ 0.0528 $ \\[0.25ex]
$ 1000 $ & TVCN & $ 0.1323 $ & $ 3.1029$ & $ 10.0394 $ & $ 0.0730$ \\[0.25ex]
 & BA & $ 0.1249 $ & $ 3.1712 $ & $ 6.7765 $ & $ 0.0386 $ \\[0.25ex]
 \hline
\end{tabular}
}
\label{tab1}
\end{table}
\vspace*{-0.5 cm}
\section{Conclusion and Future Direction}
In this paper, a time-varying network topology is designed by using preferential attachment and correlation of the nodes in the network. The network structure considers rewiring of some links of the congested node and also the removal of some anti-preferential and correlated links in the network. The correlation helps to mitigate traffic congestion in the network and provides a higher value of the critical packet generation rate, $ \lambda_c $. After that, user's data is sent through three types of shortest paths;  a path with a minimum value of $ W_g $, the shortest path with a maximum value of $ W_g $ and a randomly selected shortest path. The proposed routing approaches are applied to the network designed through different models, namely, the BA model, the TVCN model, and the proposed DTVCN model. Simulation results show that traffic capacity can be increased considerably and traffic loads are also reduced by sending data through $ W_g $(min). Moreover, the average packet travel time $ \langle T \rangle $ is reduced compared with the routing approach through $ W_g $(max) and by choosing a random shortest path (SP). Further, we analyzed various network properties and find that existence of higher betweenness centrality, $ g(max) $ and rich club (RC) phenomenon have a negative impact on the performance of the networks. 

In future work, we would like to expend our work to the more realistic environment and those can be created using network simulators.  Network congestion can be studied for varying packet generation rate on the network designed through some other strategies.

\bibliographystyle{ws-mplb}
\bibliography{mpl}
\section*{Appendix. List of Symbols}
\begin{table}[!htb]
%\caption{} % title name of the table
 % centering table
 \center
\adjustbox{max height=\dimexpr\textheight-5.5cm\relax,
           max width=\textwidth}{ 
\begin{tabular}[1]{l| p{110mm}} % creating 2 columns
\hline\hline % inserting double-line
 Symbols & Meaning
\\ [0.5ex]
\hline\hline % inserts single-line
% Entering 1st row 
$ N $ & Set of nodes \\
\hline
$ E $ & Set of links  \\
\hline
$ T $ & Life span of the networks\\
\hline
$ \lambda_i $ & Packet generation rate of node $ i $ \\
\hline
$ C_i $ & Packet forwarding rate of node $ i $ \\
\hline
$ \Pi_i^r $ & probability of a node $ i $ will be selected through preferential attachment and least correlated with its neighbor\\
\hline
$ \Pi_i^{r'} $ & probability of a node $ i $ will be selected through anti-preferential attachment and highly correlated with its neighbor\\
\hline
$ \lambda_c $ & Critical rate of packet generation \\
\hline
$ s, d $ & Source node and Destination node \\
\hline
$ g(v) $ & Betweenness centrality of a node $ v $ \\
\hline
$ e_{mn} $ & A link connects node $ m $ with $ n $ \\
\hline
$ C_{i} $ & Capacity of a node $ i $ \\
\hline
 $ \vartheta $ & Fraction of the links establish for new connections from the new node at time $ t $ , $ 0 < \vartheta < 1 $. \\
\hline
$ \gamma $ & Fraction of the links, those are rewired in the existing network, $0.5 < \gamma \leq  1 $\\
\hline
$ n_0 $ & Initial number of nodes in the seed networks.\\
\hline
$ \hat{x}(v) $ & Eigenvector centrality of node $ v $.\\
\hline
$ W_g(\sigma^r ( s \rightarrow d)) $ & Aggregate value of the BC of the shortest path between $ s $ to $ d $ for user $ r $ \\
\hline
 $ g(max) $ & A node with maximum BC \\
\hline
$ \langle T \rangle $ & Average packet travel time\\
\hline
$ \theta(\lambda) $ & An order parameter to describe the network traffic\\
\hline
$ C_lC $ & Clustering Coefficient \\
\hline
\end{tabular}
\label{tab2}
}
\end{table}

\end{document}